\documentstyle[epsfig,12pt]{article}
\begin{document}
\begin{center}
{\large{\bf The parton to hadron phase transition observed in
Pb+Pb collisions at 158 GeV per nucleon}}

\vspace{0.6 cm}
R. Stock, Fachbereich Physik, Universit\"at Frankfurt, Germany\\

\end{center}

\begin{abstract}
Hadronic yields and yield ratios observed in Pb+Pb collisions at
the SPS energy of 158 GeV per nucleon are known to resemble a
thermal equilibrium population at $T=180\pm10\:MeV$, also observed
in elementary $e^++e^-$ to hadron data at LEP. We argue that this
is the universal consequence of the QCD parton to hadron phase
transition populating the maximum entropy state. This state is
shown to survive the hadronic rescattering and expansion phase,
freezing in right after hadronization due to the very rapid
longitudinal and transverse expansion that is inferred from
Bose-Einstein pion correlation analysis of central Pb+Pb collisions.
\end{abstract}

Recent Lattice-QCD theory predicts the disappearance of the
hadronic phase of matter once the energy density exceeds a
critical value of about 1 to 1.5 GeV per $fm^3$ \cite{1}, giving
rise to a continuous, partially deconfined QCD state that is
governed by the elementary interaction of quarks and gluons. To
recreate this phase in the laboratory one collides heavy nuclei at
relativistic energy, with the goal of ascertaining the QCD
predictions, and to pin down the decay point from the partonic to
the hadronic phase by obtaining estimates for the transition
temperature and energy density. The CERN SPS Lead ($^{208}Pb)$
beam facility provides for a top energy of $158\:GeV$ per projectile
nucleon, corresponding to a total internal CM-system energy of
about $3.5\:TeV$, to heat and compress the primordial reaction
volume. In fact calorimetric data \cite{2} show that the average
transverse energy density exceeds about $2.5\:GeV/fm^3$ in that
volume, in central $Pb+Pb$ collisions. Moreover the study of $J/\Psi$
production \cite{3} demonstrates a suppression of the yield in
such collisions, characteristic of the QCD "Debeye" screening
mechanism expected in a deconfined partonic medium \cite{4}.

If one tentatively takes for granted such indications of a
deconfined state at top SPS energy one expects, likewise, to
receive signals specific to the bulk parton to hadron phase
transformation bound to occur once the primordial high density
state expands and cools toward the critical temperature and energy
density. In the present letter we shall examine the proposal
\cite{5,6,7,12} that such a signal is indeed provided for by the
composition of the bulk hadron final state, in terms of the yields
and yield ratios of the various mesonic and baryonic species. To
this end we proceed in two steps. Firstly, a synthesis is attempted
between the observation made in microscopic parton cascade and
hadronization models \cite{5,8,9} that the nonperturbative
mechanism implied in the parton to hadron "coalescence" transition
leads to a population of the hadronic species that is dominated by
conservation laws and phase space, and, on the other hand, the well
known conclusion from thermal equilibrium models \cite{7,10,11,12}
that the hadron yields resemble a hadro-chemical equilibrium
population. We shall conclude that this apparent equilibrium state
is not established by dynamical equilibration amongst the hadronic
species; it
is an outcome, and a direct fingerprint of the hadronization transition. In
the second, main step we address the question of how such a multihadronic
population pattern established right at the transition point can
survive the extensive rescattering that occurs in the course of
subsequent hadronic expansion \cite{14} in central
$Pb+Pb$ collisions. This will be answered by confronting the
dynamical expansion time scales derived from Bose-Einstein two
pion correlation analysis of such collisions \cite{15} with
estimates of the hadro-chemical relaxation time scales
\cite{16,17}, showing that the "explosive" nature of the expansion
\cite{18} prevents major rearrangement of the hadron population.

Turning first to the connection between the hadronization
mechanism and the apparent hadro-chemical equilibrium we refer to
the partonic cascade model of Ellis and Geiger \cite{8}. They
analyzed LEP data, $e^++e^-\rightarrow Z^0 \rightarrow$
hadrons at $\sqrt{s}=91\:GeV$ in a quark-gluon transport model
based on perturbative QCD. The cascade ends with quarks and gluons
recombining to form colour singlets with matching flavour
combinations in a statistical coalescence process which generates
heavy "pre-hadrons" that, in turn, decay into the final hadrons
and resonances according to their relative phase space weights. No
hadronic rescattering is considered. After resonance decay the
hadronic spectra and production multiplicities are compared to the
data finding good agreement. The authors report a second order
sensitivity of the hadron composition to the details of the
assumed parton coalescence concluding that, to first order, phase
space dominance during the hadronization phase overwhelms other
influences. The hadronic final state is thus the most probable,
maximum entropy state.

Becattini has analyzed \cite{7,10} the same LEP data using a
hadro-chemical equilibrium model based on a canonical partition
function. With fit parameters temperature $T$, reaction volume $V$
and a strangeness undersaturation parameter $\gamma_s$ he finds
good overall  agreement with the data using $T=161\:MeV$ and
$\gamma_s=0.67$. The data comprise hadron multiplicities for about
25 species ranging from $\pi$ to $\Omega$ and spread over 4 orders
of magnitude. As there is essentially no hadronic rescattering
occuring in the reaction the apparent equilibrium state can not
have originated from inelastic interactions between the 30 hadrons
created per collision event. I. e. there is no chemistry at all
{\it between} the hadrons, they form no stationary state with
dynamically maintained population ratios. In accordance with
the outcome of the Ellis-Geiger model \cite{8} we conclude that
the order seen in the hadronic population is {\it born into it} by the
non-perturbative hadronization mechanism.

Moving on to SPS $Pb+Pb$ collisions at $\sqrt{s}=17 \:GeV$ similar
data exist \cite{19} for hadron multiplicity ranging from pions to
Omegas,  and similar models are successfully applied to these
data. Geiger and Srivastava  \cite{20} have
extended the parton transport model \cite{8} to such low $\sqrt{s}$.
They report that the dominant fraction of the hadron yields (near
midrapidity) stems indeed from the parton to hadron phase
transition. They get a reasonable fit to the NA49 data \cite{19}
concerning proton, pion and kaon yields and rapidity
distributions, without any consideration of secondary hadronic
cascading: a baffling result.

Becattini et al. \cite{11} employ a generalized hadro-chemical
equilibrium model based on \cite{7} deducing a temperature
parameter of $T=192 \pm 19 \:MeV$ for the NA49 $Pb+Pb$ data
set \cite{19} which extends from pions to the
$\overline{\Lambda}/\Lambda$ ratio only. Braun-Munzinger et al.
\cite{12} address a data set for central $Pb+Pb$ combining all
available NA44, NA49 and WA97 information obtaining a satisfactory
fit at $T=170\pm11 \:MeV$. We note that the two thermal models
mentioned here differ in several ways, concerning e. g. the
treatment of excluded volume and specific approach to
strangeness saturation. The question of a final, "optimal"
formulation that avoids an inflation of free parameters is still
open \cite{11,12,13}.

We ignore this question placing prime emphasis on the
conclusion that again both the parton cascade hadronization
mechanism and the hadro-chemi\-cal equilibrium ansatz fit the data,
the latter employing "temperatures" that are slightly
{\it higher} than those obtained for elementary $e^++e^-$ collision
data. The large hadronic system created in $Pb+Pb$ through the
hadronization phase transition can thus not "cool" the hadronic
population. Although in this case the idea of re-equilibration due
to inelastic hadronic rescattering seems applicable, the above
data and results indicate that this is not occuring: the state
created at hadronization stays essentially frozen-in throughout
expansion, and its equilibrium features again appear to have
nothing at all to do with inelastic interactions among the 2500
hadrons \cite{19} in the final expansion. Solving this puzzle will
help to establish the multi-hadronic population ratios as a
fingerprint of a parton to hadron phase transition occuring in SPS
collisions.

Hadronic inelasticity is rather weak in a hadronic medium at
$T=180\:MeV$: the mean CM kinetic energy of a pair of baryons is only
$270\:MeV$ and we recall that the average pion multiplicity in $p+p$
collisions at this energy \cite{21} is only about 0.2. Only every
fifth collision is inelastic.
Hadronic equilibration times should thus be well above $1\:fm/c$
at this temperature,  growing to about $20\:fm/c$ at $T=120\:MeV$
\cite{16}, the temperature of
final hadronic decoupling from strong interaction \cite{15}.

To appreciate these relaxation time scales we need to know about the expansion
time scale in central $Pb+Pb$ collisions. Fig.1 shows the time
profile of pion decoupling \cite{22} from the fireball. The
Gaussian with mean $\tau_f=8\:fm/c$ and width $\Delta \tau=4\:fm/c$
results from an analysis of Bose-Einstein pion pair correlations
in central $Pb+Pb$ by NA49 \cite{15} using a dynamical hadron
source expansion model developed by Heinz \cite{23}. The time scale
in this model begins after the primordial source is formed,
irrespective of its content at this time (partons or hadrons).
This time is reached at complete interpenetration,
$t \approx R(Pb) \gamma^{-1}_{CM} \approx 1 \:fm/c$. Then the
emission clock starts ($\tau=0$ in Fig. 1), correctly so as pions
can be emitted from the surface of the fireball right away. We see
that the mean decoupling (freeze-out) time of pions is $\tau=\tau_f=8\:fm/c$
from this analysis. This includes the average life-time of the
partonic phase. In order to arrive at an estimate of the mean
hadronization time we reproduce  in Fig. 2 a result of the
partonic cascade model by
Geiger and Srivastava  \cite{20} that we quoted above to reproduce
global features of central $Pb+Pb$ collisions. This shows the time
dependence of the average energy density in the vicinity of
midrapidity as exhibited by partons and hadrons (plus resonances),
 in a cylindrical
subvolume of $4\:fm$ transverse radius, i. e. in the interior
section of the source. The partonic era ends at $t=6\:fm/c$. We
may crudely estimate a mean hadronization time of $3\:fm/c$.

The average time span between hadron formation and freeze-out may
thus be estimated to be about $6\:fm/c$ (taking proper account of
the fact that $\tau=0$ in Fig.1 corresponds to $t=1\:fm/c$ in Fig.2).
However, during this time interval any freshly formed hadron
experiences a dramatic falloff concerning density and temperature
in its local, co-moving environment cell that participates in the
overall expansion of the system. A closer look at Fig.2 already
reveals the presence of this expansion: the (energy) density of
hadrons and resonances in the inner subvolume
considered here stays constant throughout the hadronization phase
although it is constantly replentished by further, newly created
hadrons. This implies that the rate of hadronization inside the
cylinder must be compensated by the rate of hadrons escaping,
thus preventing a pile up to unreasonable central
density beyond $0.6 \:GeV/fm^3$. A more quantitative consideration
shows that the newly
formed hadrons must have left this inner subvolume of
maximal density and temperature after about $1.5\:fm/c$ on
average, much smaller than the estimated typical hadronic
relaxation time  of about $4\:fm/c$, that was obtained by Mekjian
and Kapusta \cite{16} for kaons in a hadron gas at $T=180 \:MeV$.
Travelling outward in
longitudinal and (more slowly) in transverse direction the average
hadron encounters a rapidly diluting and cooling environment. In
fact a combined analysis of pion pair correlation and hadron
spectral data for central $Pb+Pb$ SPS collisions \cite{15} has
revealed evidence for collective outward velocity fields reaching
up to $\beta_{\bot} \approx 0.55$ (reminiscent of the limiting
hydrodynamical fluid velocity of $c/\sqrt{3}$), the temperature
falling to $T=120\:MeV$ at hadronic decoupling. This fall thus
occurs during an average time span of $6\:fm/c$, during which
the total reaction volume increases about tenfold \cite{12,15}.
 While the average
hadron encounters about 6 binary collisions \cite{14}, perhaps
maintaining local momentum equilibrium, the inelastic
fraction fades away rapidly in such an "explosive" hadronic
expansion \cite{18}. Thus the primordial population pattern
created by the hadronization transition among hadronic and
resonance species stays essentially unaltered (the resonances
decaying within their proper time), remaining "frozen in"
throughout the hadronic expansion phase.

The fingerprint of the parton to hadron phase transformation in
central $Pb+Pb$ collisions is thus preserved in the observed bulk
hadron yields, and yield ratios. Moreover the models of
microscopic parton transport toward the critical QCD conditions of
hadronization indicate statistical phase space dominance,
according to the spin, mass and flavour spectrum of created
hadronic and resonance states. Not surprisingly, the thermal
hadrochemistry models \cite{11,12,13} thus come up with meaningful
accounts of this initial maximum entropy state. This lifts the
conceptual qualms ever prevailing since Hagedorns engagement \cite{24}
with statistical thermodynamics applied to hadron production data.
There is no chemical equilibration occuring at the
hadronic level to dynamically {\it achieve} a state of limiting
temperature, neither in elementary $e^++e^-$ or $p+\overline{p}$
collisions nor in ultra-relativistic nuclear collisions \cite{25}.
Such states are created from partonic energy densities and
temperatures {\it above} the limiting hadronic conditions,
the QCD hadronization thus populating a statistically ordered
system (to avoid the term "state" as it invites the view of
stationary conditions) just at the limiting hadronic energy
density. This results in the apparent universality \cite{7} of
hadron production  as reflected in thermal model analysis
from $e^++e^-$ to $Pb+Pb$.

In summary
once we take for granted the tentative evidence from $J/\Psi$
suppression, and from  estimates of primordial energy density,
in favour of the hypothesis that SPS $Pb+Pb$ collisions reach
beyond the limit of the hadronic phase of matter (as predicted
by lattice QCD theory) the system has to re-hadronize upon
expansion. And, second, the properties of this long-sought
parton to hadron phase transition are reflected in
characteristic features of the bulk hadronic
phase that survive the rescattering and expansion phase because
they freeze-in right at, or near the phase transition. Our
focus was placed on the regularities of hadronic yield ratios
because they are allowing for an estimate of the critical
transition temperature (and the corresponding value of the
baryo-chemical potential $\mu_B$ that we did not comment about) owing to
the applicability of thermal model analysis to this maximum entropy
state that we have tried to establish. One now expects a
convergence among the various thermal model approaches \cite{11,12,13},
to finally pin down $T$ and $\mu_B$ of the phase transition.
At present we may conclude that the QCD transition temperature
$T_c=170-180 \:MeV$, at $\mu_B \approx 0.25 \:GeV$.

A final question requires attention: does universality of hadronization
imply that small and large systems hadronize by the identical same
QCD mechanism?  The
author is not aware of a rigorous QCD argument. The
colour and flavour composition of the partonic phase should
reflect in the hadronization outcome (if there are mostly gluons,
if strange quarks are far underpopulated or near-equilibrated etc.).
Moreover there are
rather straight forward effects of increasing system size. In
the phenomenological coalescence approach Ellis and Geiger
\cite{8} have studied the step from $e^++e^- \rightarrow Z^0 \rightarrow$
hadrons to $e^++e^- \rightarrow W^+ W^- \rightarrow$ hadrons (at
LEP 2). In the latter case, the two parton cascades may overlap.
This merging of the decaying subvolumes  leads to total multiplicity
increases, and the transverse momentum spectra
exhibit some softening. The LEP 2 data have not been analyzed yet
to demonstrate such effects. However, comparison of elementary $p+p$
data with central $Pb+Pb$ collisions shows a drastic effect in the
hadronic strangeness composition: the ratio of strange to nonstrange quarks
contained in the hadrons increases by a factor 2 \cite{11,19}, the
ratio of double strange cascade hyperons to nucleons by a factor
of about 10 \cite{26}, the same ratio for $\Omega$ hyperons by
a  factor approaching 20 \cite{27}. Thus there is no flavour scaling
from the elementary nucleon-nucleon to central $Pb+Pb$ collisions.
A certain fraction of this effect has been explained in the
hadro-chemical model \cite{11,28} to arise from the so-called
"canonical suppression" (e. g. of strangeness) dominating the small
system, due to local difficulty of energy and quantum number
conservation \cite {29}. This effect disappears with increasing system
size, a precursor seen in the overlapping $W^+W^-$ cascade decay
mode \cite{8}, and plays no role in central $Pb+Pb$ collisions which thus
receive a grand-canonical treatment in these models. However, there is still
further strangeness enhancement beyond this mechanism, which may be
traced to before the phase transition assuming a higher relative
strangeness content in the partonic $Pb+Pb$ collision phase than
in $p+p$ collisions \cite{28}. Primordial flavour or gluonic
composition levels, perhaps characteristic of a symmetric
partonic phase approaching the ideal QCD plasma state \cite{28}
may thus survive even the processes occuring during the hadronization phase
transition \cite{30}. This then reflects in additional (e. g. flavour) parameters
in the hadrochemical model. This new aspect (see, however, ref.
\cite{5}) adds further significance to bulk hadronic signals from
SPS nucleus-nucleus collisions.

\begin{figure}
\hspace{3cm}
\epsfig{figure=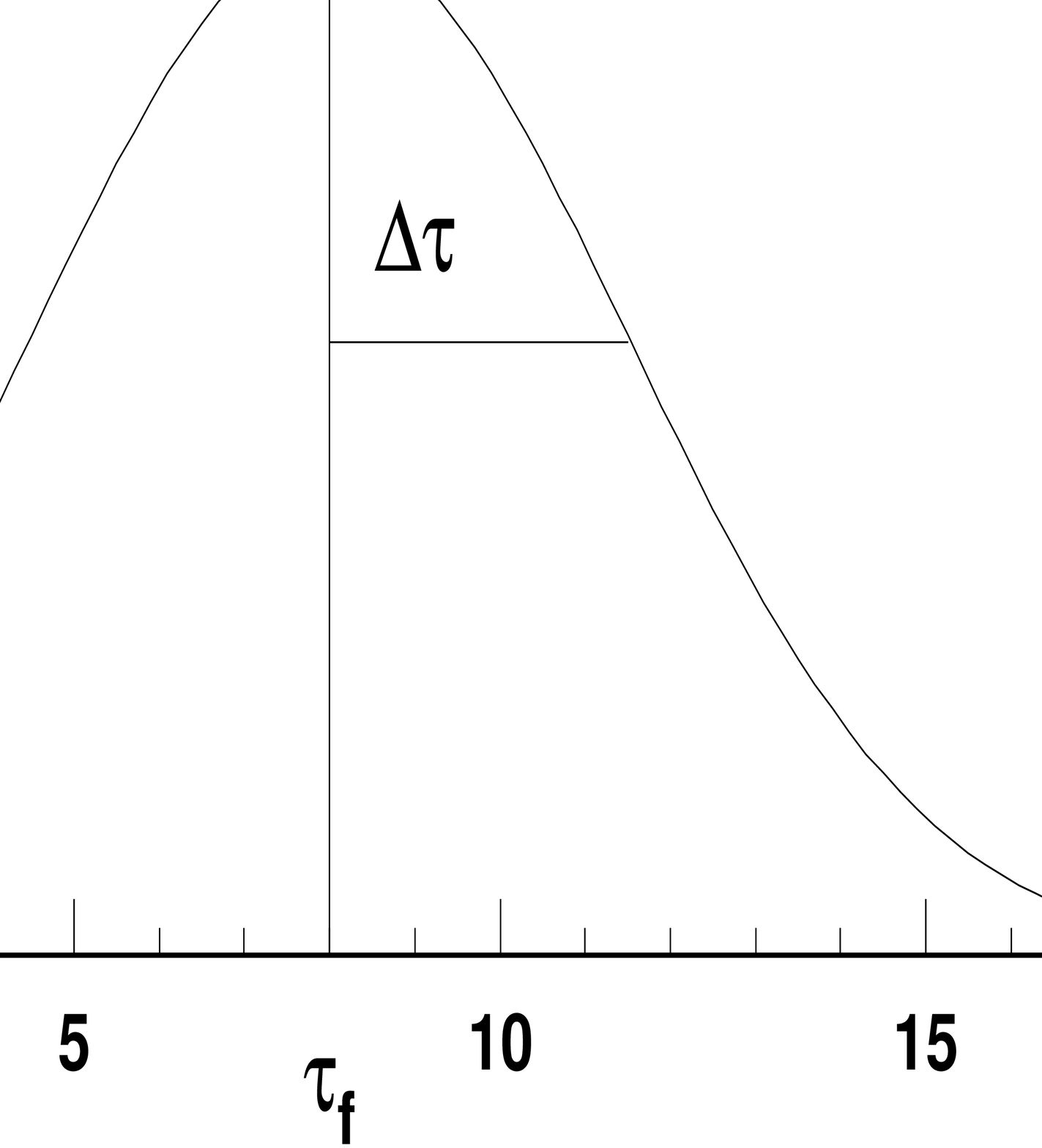,height=8cm, width=7cm}
\caption{Time profile of the pion decoupling strength from the
hadronic fireball formed in central $Pb+Pb$ collisions
(arbitrarily normalized to unity). From two pion correlation
analysis \cite{15,22,23} we get an average decoupling time
$<\tau_f>=8 \:fm/c$, and a mean duration of the decoupling phase
characterized by a width $\Delta\tau=4 \:fm/c$
for the assumed Gaussian profile.}
\end{figure}

\newpage

\begin{figure}
\epsfig{figure=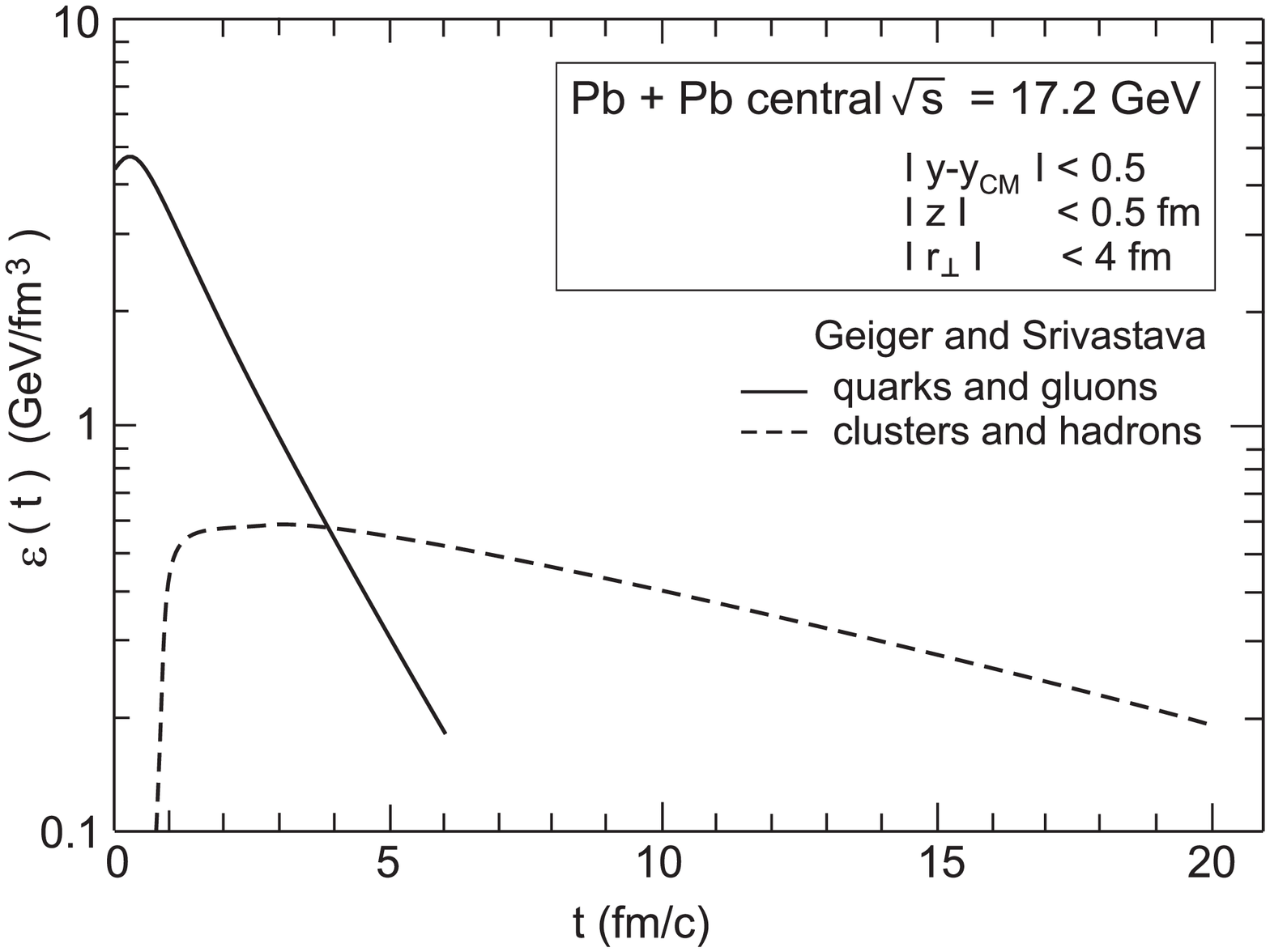, angle=270, width=13cm}
\vspace{1cm}
\caption{Parton cascade model prediction for the time profile of the energy
density contained in partons and hadrons, as found in a central
cylinder subvolume of $4 \: fm/c$ transverse radius at $|y-y_{CM}|<0.5$
in central $Pb+Pb$ collisions at $\sqrt{s}=17.2 \:GeV$. Adapted
from Geiger and Srivastava \cite{20}.}
\end{figure}


\begin{thebibliography}{99}
\bibitem{25} Note however that at modestly relativistic energy
(about $1\:GeV$ per nucleon) central collisions of heavy nuclei
at Bevalac and SIS may indeed reach a genuine dynamical chemical
equilibrium governed by approximately equal forward and backward
reaction rates \cite{17}. But these near-equilibrium states last for up to
$10\:fm/c$ and  comprise essentially nucleons and deltas
only, staying far from the limiting temperature.
\end{thebibliography}
\end{document}